\documentstyle[aps]{revtex}
\begin{document}
\newcommand{\beq}{\begin{equation}}
\newcommand{\eeq}{\end{equation}}
\newcommand{\beqn}{\begin{eqnarray}}
\newcommand{\eeqn}{\end{eqnarray}}
\newcommand{\bmath}{\begin{mathletters}}
\newcommand{\emath}{\end{mathletters}}
\twocolumn[\hsize\textwidth\columnwidth\hsize\csname @twocolumnfalse\endcsname 
\title{Superconductivity from Hole Undressing}
\author{J. E. Hirsch }
\address{Department of Physics, University of California, San Diego\\
La Jolla, CA 92093-0319}
 
\date{\today} 
\maketitle 
\begin{abstract}
Photoemission and optical experiments indicate that the transition to   
superconductivity in cuprates is an 'undressing' transition .                 
 In photoemission this is seen as a coherent                        
quasiparticle peak emerging from an incoherent background, in optics as   
violation of the Ferrell-Glover-Tinkham sum rule indicating effective mass 
reduction of superconducting carriers. We propose that this is a 
manifestation of the fundamental electron-hole          
asymmetry of condensed matter described by the theory of hole superconductivity.
The theory asserts that electrons in nearly empty bands and holes in nearly full bands 
are fundamentally different : the former yield
high conductivity and normal metals, the latter yield
low normal state conductivity and high temperature superconductivity.
This is because the normal state transport of electrons is coherent and
that of holes is incoherent. We explain how this asymmetry arises from
the Coulomb interaction between electrons in atoms and the nature of atomic
orbitals, and propose a simple Hamiltonian to describe it. A $universal$ mechanism 
for superconductivity follows from this physics, whereby dressed hole carriers 
undress by pairing, turning (partially) into electrons
and becoming more mobile in the superconducting state. 

\end{abstract}
\vskip2pc]

Holes are not like electrons\cite{web,hole}. This fundamental fact continues
to be ignored in contemporary solid state science\cite{heis}.
By holes and electrons we mean the charge carriers in 
energy bands in a metal when the Fermi level is near the top and near the bottom 
of the band respectively.

The physical difference between electron and hole carriers is not captured by
Hamiltonians commonly used to study many-body phenomena in solids such as
the Hubbard model. Thus, new Hamiltonians need to be considered that 
contain the fundamental charge asymmetry of condensed matter. The generalized
Hubbard model with correlated hopping\cite{camp} contains part of the
essential physics of electron-hole asymmetry. The other part is incorporated 
by Hamiltonians that describe also high energy degrees of freedom, which may
be generically termed electron-hole asymmetric polaron models\cite{polaron,undr}.
These models can be formulated with only electronic degrees of freedom\cite{elec}
(more than one band is needed) or with coupled electronic and bosonic degrees
of freedom\cite{spin,holst,undr}. The generalized Hubbard model with correlated hopping
is the low energy effective Hamiltonian that arises from these more fundamental
models.

We argue that the transport of electricity is fundamentally different when the carriers 
in the metal 
are electrons and when they are holes. Holes cause a large disruption in their
electronic environment when they propagate, while electrons cause little or no 
disruption. As a consequence, holes are heavy and electrons are light, $and$,
hole transport is largely incoherent while electron transport is largely coherent.
The qualitative difference in the frequency dependent conductivity of electrons
and holes is depicted in Figure 1. As the number of physical electrons in a 
nearly empty band is increased, electrons turn gradually into holes, and
become heavier and more incoherent. This is because they become increasingly
'dressed', due to electron-electron interactions, as depicted schematically
in figure 2. Conversely, as the number of physical electrons in a nearly full
band is decreased, holes turn gradually into electrons, and become lighter
and more coherent as they gradually undress. The other way that hole carriers
can turn into electrons is by pairing, thereby mimicking locally
a situation where the band is less full. As a consequence,
superconductivity will be characterized by the carriers becoming lighter and
more coherent in the superconducting than in the normal state, and kinetic
energy will be lowered\cite{kinetic}. Since pairing of electron carriers
would instead lead to carriers becoming heavier and more incoherent in
the superconducting state, and increased kinetic energy, pairing of
electron carriers $cannot$ $drive$ $superconductivity$.

A single electron in an empty band is a non-interacting particle, with
spectral function given by
\beq
A(k,\omega)=\delta(\omega-\epsilon_k)
\eeq
with $\epsilon_k$ the band energy. The single particle spectral function in
a many-body system, for momentum $k$ close to the Fermi surface, is of
the form
\beq
A(k,\omega)=z\delta(\omega-\epsilon_k)+A'(k,\omega)
\eeq
where $z$ is the quasiparticle weight, and $A'$ is the incoherent part of
the spectral function (we have neglected possible dependence of $z$ on $k$).
One has in general $z<1$, and only for non-interacting particles is
$z=1$. The normalization condition $\int A(k,\omega)d\omega=1$ ensures that
$A'$ will be nonzero whenever $z<1$. A single hole in a full band has $z<1$ and
$A'\neq 0$, thus is qualitatively different from a single electron
in an empty band (Eq. (1)) for which $z=1$ and $A'=0$.

The difference between a single electron in an empty band and a single
hole in a full band is simplest to understand in a tight binding formulation.
Consider as the simplest case the band formed by overlap of $1s$ orbitals in a
lattice of hydrogen-like ions, with nuclear charge $Z$. The process of creating
an electron in the empty $1s$ orbital (figure 3a)
\beq
|\uparrow >=\varphi_{1s}(r)=Ce^{-Zr/a_0}
\eeq
($a_0=$Bohr radius) does not affect any other degree of freedom. Hence the
single site spectral function is given by
\beq
A_0(\omega)=\delta(\omega-\epsilon_{1s})
\eeq
(with $\epsilon_{1s}=-13.6eV$). Consider instead the process of creating
an electron of spin $\uparrow$ when an electron of spin $\downarrow$ already
exists in the $1s$ orbital (figure 3b). If the second electron is created in the
$1s$ orbital also, a state of very high energy will result, due to the
large Coulomb repulsion between two electrons in that orbital ($17Z eV$).
Instead, consider the ground state of the two-electron ion in the Hartree
approximation
\bmath
\beq
|\uparrow \downarrow ^0>=\bar{\varphi}_{1s}(r_1)\bar{\varphi}_{1s}(r_2)
\eeq
\beq
\bar{\varphi}_{1s}(r)=\bar{C}e^{-\bar{Z}r/a_0}
\eeq
\beq
\bar{Z}=Z-5/16
\eeq
\emath
To obtain the lowest energy state we want to create the second electron
in the expanded orbital $\bar{\varphi}_{1s}$, with creation 
operator $\bar{c}^\dagger_\uparrow $.
We have then
\beq
\bar{c}^\dagger_\uparrow |\downarrow >=|\uparrow \downarrow ^0>
<\uparrow \downarrow ^0| \bar{c}^\dagger_\uparrow |\downarrow >+
\sum_{l\neq 0}  |\uparrow \downarrow ^l>
<\uparrow \downarrow ^l| \bar{c}^\dagger_\uparrow |\downarrow >
\eeq
where $|\uparrow \downarrow ^l>$ are a complete set of excited states of
the doubly occupied ion, with energies $\bar{\epsilon}_{1s}^{(l)}$. The
single particle spectral function for this process is
\bmath
\beq
A_1(\omega)=z_h\delta(\omega-(\bar{\epsilon}_{1s}^{(0)}-\epsilon_{1s}) )+A'_1(\omega)
\eeq
\beq
z_h=|<\uparrow \downarrow ^0| \bar{c}^\dagger_\uparrow |\downarrow >|^2
\eeq
\beq
A'_1(\omega)=\sum_{l\neq 0} <\uparrow \downarrow ^l| \bar{c}^\dagger_\uparrow |\downarrow >|^2
\delta(\omega-(\bar{\epsilon}_{1s}^{(l)}-\epsilon_{1s}))
\eeq
\emath
hence it has now an incoherent part, and a quasiparticle weight $z_h<1$.
This is because when the $\uparrow$ electron is created the pre-existing
$\downarrow$ electron will change its state, from $\varphi_{1s}$ to the
self-consistent state $\bar{\varphi}_{1s}$; in addition, the
end result of this process may be any of the excited states of the doubly
occupied ion, as given by Eq. (6). Similarly the spectral function for
creating a hole (destroying an electron) in the doubly occupied ion is
\bmath
\beq
A_2(\omega)=z_h\delta(\omega+(\epsilon_{1s}-\bar{\epsilon}_{1s}^{(0)})) +A'_2(\omega)
\eeq
\beq
A'_2(\omega)=\sum_{l\neq 0} <\uparrow \downarrow ^l| \bar{c}^\dagger_\uparrow |\downarrow >|^2
\delta(\omega+(\epsilon_{1s}^{(l)}-\bar{\epsilon}_{1s^{(0)}}))
\eeq
\emath

Eqs. (4) and (7), (8) show the fundamental difference between the spectral
functions for creating electrons in empty bands and creating or destroying holes
in full bands.  The qualitative structure does
not depend on the Hartree approximation, and illustrates the general
fact that the hole spectral function will have a large incoherent
contribution and a small quasiparticle weight, as shown schematically
in Figure 4. This is due to the physical fact that the spacing
between atomic energy levels is smaller than the Coulomb repulsion 
between electrons in a level, thus when an electron is created in an
already occupied level the wavefunctions will expand into other
atomic levels. This physics $cannot$ be captured with models with a single
orbital per site\cite{inapp}. This effect then leads to the fundamental difference in
the transport properties when the band is almost empty and when it is
almost full. For electrons in nearly empty bands the hopping between
atoms $t$ is unrenormalized, while for holes in a full band it is 
reduced by the quasiparticle weight
\beq
t_h=z_h t
\eeq
due to the disruption caused in the other electron in the orbital during the
hopping process, as depicted schematically in figure 5. This leads to an 
enhanced effective mass for holes, hence a larger dc resistivity,
and to a large incoherent contribution to the frequency dependent
conductivity from hopping processes where the electrons end up in
excited states.

We can find the quasiparticle weight for the hole explicitely in this
Hartree approximation
\beq
z_h=|<\varphi_{1s}|\bar{\varphi}_{1s}>|^2=
\frac{(1-\frac{5}{16Z})^3}{(1-\frac{5}{32Z})^6}
\eeq
which approaches unity as the nuclear charge $Z$ increases, and becomes
small as $Z\rightarrow 0.3125$. Even though obtained in the Hartree
approximation, the qualitative effect will be generally true
(and even larger when one goes beyond the Hartree approximation\cite{elhole2}).
When the effective nuclear charge $Z$ is large $z_h\rightarrow 1$, hence
hole quasiparticles become coherent and light, and resemble electron
quasiparticles. Instead, when the effective nuclear charge is small, 
$z_h\rightarrow 0$ and holes
become very heavy and incoherent. This is the regime most favorable
for high temperature superconductivity. Note how this is in qualitative agreement with the 
situation in cuprates, where the relevant ions, $O^=$, are highly
negatively charged, corresponding to a small $Z$. Moreover the 
entire $CuO$ planes are highly negatively charged, giving rise to 
highly 'floppy' orbitals thus creating the most favorable environment for
hole pairing. This also provides a rationale for the advantage of the
two-dimensional
structure. Since the negative charge needs to be compensated for charge
neutrality, one way to do it is to arrange the excess negative charge
in the conducting two-dimensional planes where the pairing occurs and the
compensating positive charge outside the main conducting structures.
It is possible that highly negatively charged conducting substructures
could also be created in three-dimensional structures, which would be even
more advantageous for high temperature superconductivity since, everything
else being equal, higher coordination strongly enhances superconductivity
in this model\cite{holetheory}.

Superconductivity occurs in this theory from the enhanced hopping
amplitude when two hole carriers pair. In that case the hopping is
$t_h'=\sqrt{z_h} t$ and the difference $\Delta t=t_h'-t_h$ drives the transition to
superconductivity, as discussed in detail elsewhere\cite{holetheory}.
The superconducting condensation energy is kinetic\cite{kinetic},
and the quasiparticle weight is larger in the superconducting than
in the normal state\cite{undr}.

To study quantitatively the physics described above requires information on
excitation energies and matrix elements of electronic states in
multi-electron atoms, and is a difficult many-body problem. It is
useful to first understand throughly the novel physical phenomena that
emerge, for which we can use a variety of model Hamiltonians that 
contain the essential physics\cite{undr,elec,spin,holst}. As perhaps the simplest realization,
consider the site Hamiltonian
\beq
H_i=\omega_0a_i^\dagger a_i+g\omega_0(a_i^\dagger+a_i)
n_{i\uparrow }n_{i\downarrow }
\eeq
where $a_i$ is a local boson operator and $n_{i\sigma}$ is an
$electron$ number operator. This is a special case of the
generalized Holstein model discussed in Ref. 9, with fully non-linear coupling.
Using a generalized Lang-Firsov transformation the following relation
between electron particle ($c_{i\sigma}^\dagger$) and quasiparticle 
($\tilde{c}_{i\sigma}^\dagger$) operators results\cite{undr}:
\beq
c_{i\sigma}^\dagger=e^{g(a_i^\dagger-a_i)\tilde{n}_{i,-\sigma}}\tilde{c}_{i\sigma}^\dagger 
\eeq
so that for an empty site, creating an electron particle is the
same as creating a quasiparticle
\bmath
\beq
c_{i\sigma}^\dagger=\tilde{c}_{i\sigma}^\dagger
\eeq
and the quasiparticle weight is $1$, while instead if the site is already occupied 
by an electron of opposite spin,
\beq
c_{i\sigma}^\dagger=e^{g(a_i^\dagger-a_i)}\tilde{c}_{i\sigma}^\dagger .
\eeq
\emath
Taking the ground state expectation value of the exponential in Eq. (12)
yields
\beq
<e^{g(a_i^\dagger-a_i)\tilde{n}_{i,-\sigma}}>=e^{-(g^2/2)\tilde{n}_{i,-\sigma}}
\eeq
so that the hole quasiparticle spectral weight is
\beq
z_h=e^{-g^2}
\eeq
and holes become heavily dressed if $g$ is large, while electrons
remain undressed. The quasiparticle weight for general electronic band
filling $n$ is given by
\beq
z(n)=[1+\frac{n}{2}(e^{-g^2/2}-1)]^2
\eeq
which interpolates between $1$ and $z_h$ and quantifies the magnitude of coherent 
response of the system for given band filling.
The relation Eq. (12) can be written as
\beq
c_{i\sigma}^\dagger=[1+(e^{-g^2/2}-1)\tilde{n_{i,-\sigma}}]\tilde{c}_{i\sigma}^\dagger
+\tilde{n}_{i,-\sigma}\times incoherent \ part
\eeq
where the 'incoherent part' describes the processes where bosons are created
when the electron is created at the site. This represents the
second term in Eq. (6), where the ion ends up in an excited state when the
second electron is created. The full Hamiltonian to be studied is then
\beq
H=\sum_i H_i-\sum_{<ij>}t_{ij} (
c_{i\sigma}^\dagger c_{j\sigma}+h.c.) + \sum_{<ij>}V_{ij}n_i n_j
\eeq
with $H_i$ given by Eq. (11). The low energy effective Hamiltonian
for quasiparticles that results, using Eq. (14), is
\bmath
\beq
H_{eff}=-\sum_{<ij>}t_{ij}^\sigma (
\tilde{c}_{i\sigma}^\dagger \tilde{c}_{j\sigma}+h.c.) + 
\sum_{<ij>}V_{ij}\tilde{n}_i \tilde{n}_j
\eeq
\beq
t_{ij}^\sigma=t_{ij}[1-(1-\sqrt{z_h})(\tilde{n}_{i,-\sigma}+\tilde{n}_{j,-\sigma})
+(1-\sqrt{z_h})^2 \tilde{n}_{i,-\sigma} \tilde{n}_{j,-\sigma}]
\eeq
\emath
which in particular describes the superconducting state\cite{holetheory}. However to
understand the fundamental processes of spectral weight transfer
that occur it is necessary to use the full Hamiltonian Eq. (17)\cite{polaron,undr} . This
Hamiltonian predicts that spectral weight will be transfered from the
high energy scale determined by $\omega_0$, which represents an electronic
excitation energy scale of the multielectron atom, down to low (intraband)
energies, both when the system goes superconducting and when it is
doped with holes, both in the one-particle spectral function
(photoemission\cite{dingfeng}) and in the two-particle spectral
function (optical conductivity\cite{uchfugbas}).

We believe that the physics described by these models represents a new
paradigm in many-body physics. When the system goes superconducting,
hole quasiparticles 'undress' and become more like bare particles.
In the conventional Fermi liquid approach quasiparticles are fixed
objects, that interact weakly with one another and develop special
correlations when a transition to a collective state occurs, but do not change
their intrinsic nature. Here instead quasiparticles  change
their most fundamental properties, their weight and their effective mass,
as the ordered state develops. The energy that drives the transition to 
superconductivity originates in the very large energy renormalization
that is involved in going from a description based on bare, strongly interacting
particles,  to one based on dressed, weakly interacting quasiparticles.
The theory of hole superconductivity proposes that holes, by undressing
and turning into electrons, manage to take advantage of 
this rich source of untapped energy.

What are measurable consequences of the fundamental charge asymmetry on which
this theory is based? We have shown that it leads to universal asymmetry in
NIS tunneling\cite{tunnel}, with larger conductance predicted for a negatively
biased sample, and to the prediction of positive thermoelectric power for
NIS and SIS tunnel junctions\cite{thermo}. There exists some experimental evidence for
the former, while the latter has not been experimentally tested. Furthermore,
we have recently proposed that it should lead to negatively charged vortices
in the mixed state of type II superconductors, and more generally to a tendency
for superconductors to expel negative charge from the bulk\cite{charge}, leading to higher 
negative charge density and superfluid density near the surface, as shown
schematically in figure 6. Because the principles on which the theory is
based are very general, we expect that if the theory is valid it should apply
to $all$ superconducting materials\cite{other}.

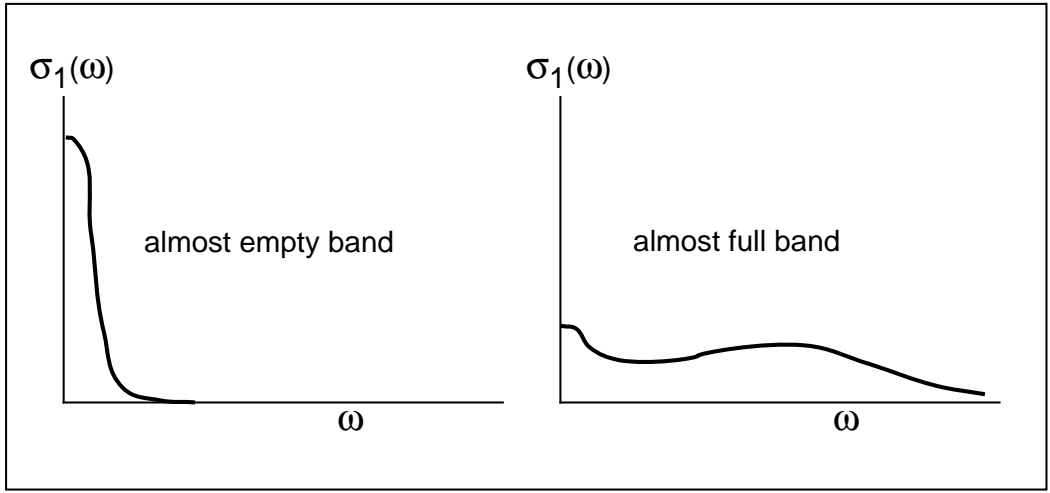
\begin{figure}
\caption { 
Qualitative difference between frequency-dependent conductivity when
the conduction band is almost empty (electron carriers) and when it is
almost full(hole carriers) (schematic). Holes have a large effective
mass, resulting in a small value of $\sigma_1(\omega=0)$, and give rise
to a large incoherent contribution to $\sigma_1(\omega)$ extending to
frequencies well beyond the band energies. Electrons give rise
to Drude-like conductivity with large $\sigma_1(\omega=0)$ due to their small
effective mass.
}
\label{Fig. 1}
\end{figure}

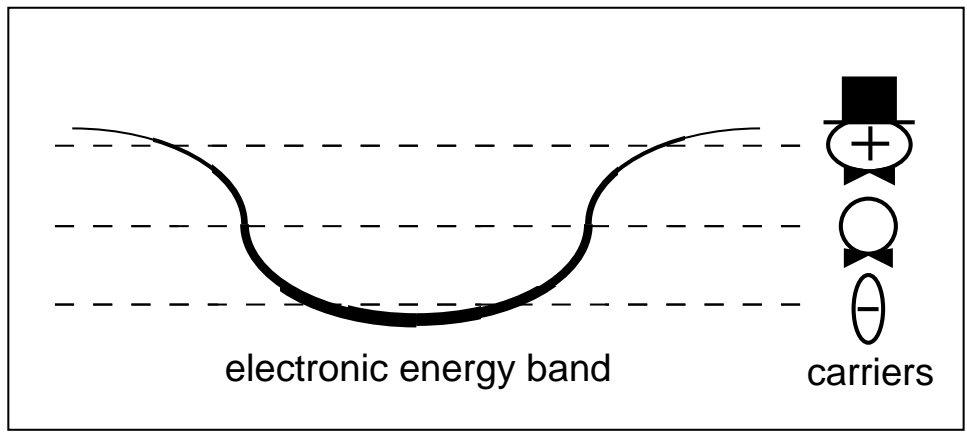
\begin{figure}
\caption { 
When the Fermi level is near the bottom of a band, carriers are 
undressed, light, coherent, and electron-like. As electrons are added to
the band and the Fermi level rises, carriers become increasingly
dressed, heavier, incoherent, and hole-like. The thickness of the
$\epsilon$ versus $k$ line indicates qualitatively the strength of 
the quasiparticle weight.
}
\label{Fig. 2}
\end{figure}

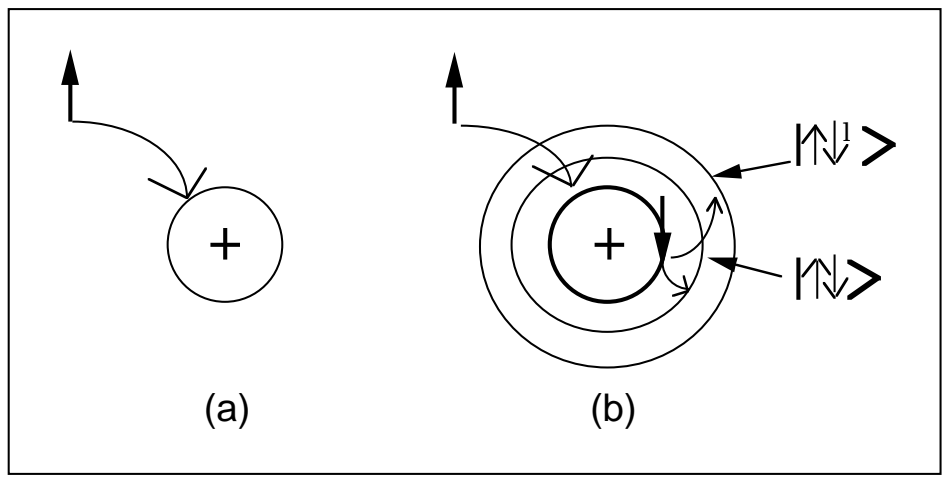
\begin{figure}
\caption {An electron created in the empty $1s$ orbital of the ion (a) 
produces no disruption in another degree of freedom; an electron created
in the orbital that is already occupied by another electron (b) will creat
a disruption of that degree of freedom, and the resulting state can be 
any of the excited states of the doubly occupied ion. Similarly a hole
created in the doubly occupied ion (an electro destroyed) will leave
the remaining electron in any of the possible excited states of
the singly occupied ion.
}
\label{Fig. 3}
\end{figure}

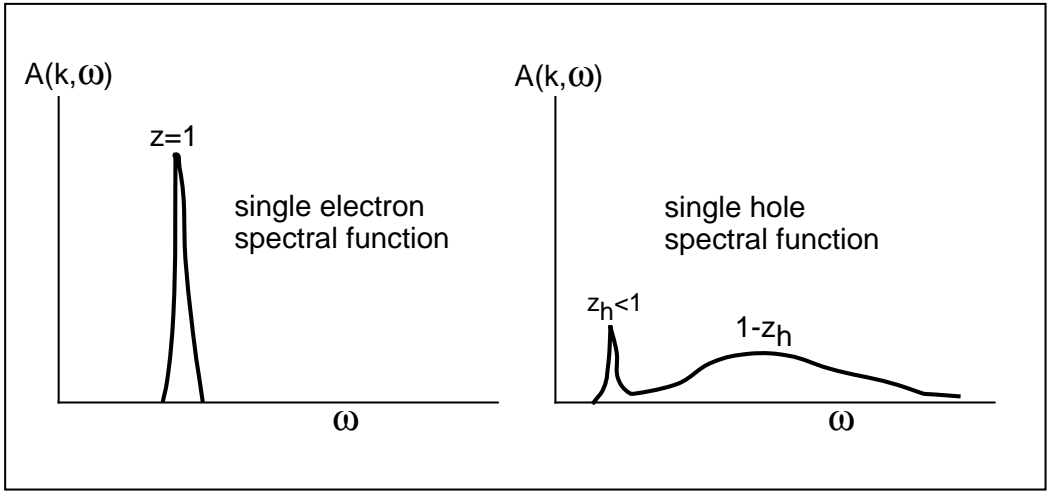
\begin{figure}
\caption { 
Qualitative difference of single particle spectral functions
for particles near the Fermi level for a nearly empty band (electrons) and a 
nearly full band (holes). For the nearly full band the quasiparticle
spectral weight is small and a large incoherent contribution exists. For the nearly
empty band the spectral function is entirely coherent.
}
\label{Fig. 4}
\end{figure}

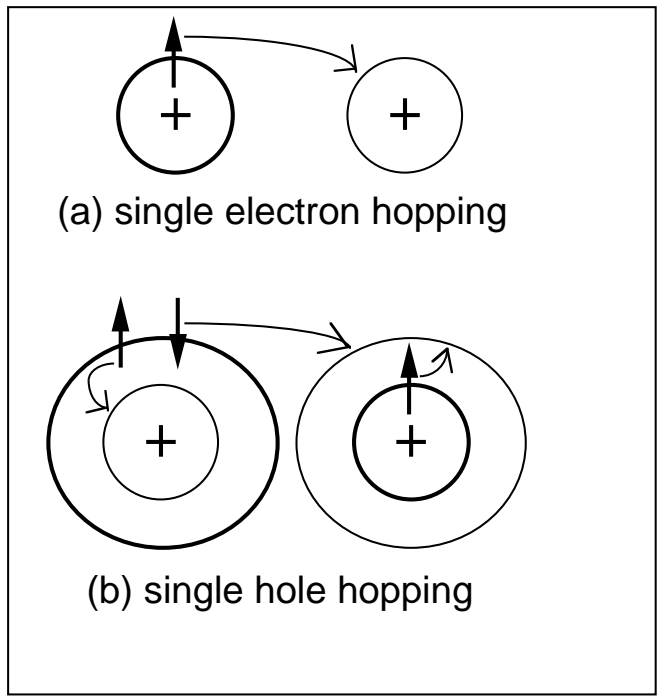
\begin{figure}
\caption { 
Hopping processes giving rise to conduction when the band is almost
empty (a) and when it is almost full (b). In (b) the 'diagonal
hopping processes', where the ions make ground state to ground state
transitions, involve a rearrangement of the electrons that are not
hopping. In addition, the hopping process may lead to the ion
making a transition to an excited state.
}
\label{Fig. 5}
\end{figure}

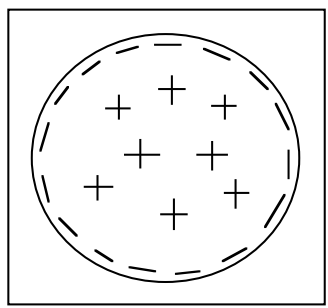
\begin{figure}
\caption { 
Schematic picture of a spherical superconducting body. Negative charge is
expelled from the bulk to the surface.
}
\label{Fig. 6}
\end{figure}

\end{document}